\documentclass{ws-procs9x6}

\begin{document}

\title{Recent progress towards a chiral effective field theory for the NN
system}
\author{C.J. Yang$^*$ and Bingwei Long}
\address{Dipartimento di Fisica, Universita di Trento, via Sommarive, 14 I-38123
Trento, Italy \\
$^*$E-mail: chieh@science.unitn.it
}

\address{Department of Physics, Sichuan University, 29 Wang-Jiang Road, Chengdu,
Sichuan 610064, China\\
E-mail: bingwei@scu.edu.cn}

\begin{abstract}
Since Weinberg's proposal two decades ago, chiral effective field theory in
the NN sector has been developed and applied up to order $O((Q/M_{hi})^4)$.
In principle it could provide a model-independent description of nuclear
force from QCD. However, in spite of its huge success, some open issues such
as the renormalization group invariance and power counting, still remain to
be solved. In this talk we refine the chiral effective field theory approach
to the NN system based on a renormalization group analysis. Our results show
that a truly model-independent description of NN system can be obtained by a
new power counting which treats the subleading order corrections
perturbatively.
\end{abstract}


\bodymatter


\section{Introduction}

\label{sec:introduction}

Based on the symmetries of Quantum chromodynamics (QCD) in the low energy
region ($\leq $ 1 GeV), chiral effective field theory ($\chi $EFT) enables
calculations of strong interaction in the non-perturbative region. However,
unlike the pion-pion and pion-nulceon section, where the power
counting---the key ingredient which guarantees the intrinsic consistency of
an EFT--- is given clearly from the vertices generated by the chiral
Lagrangian, the power counting in nucleon-nucleon (NN) case is hindered by
the infrared enhancement and cannot be obtained straightforwardly.

The first step out of the NN problem, as suggested by Weinberg\cite{We90},
is to apply the power counting to the NN potential level first, and then sum
the amplitude by iterating the potential in Schrodinger or
Lippmann-Schwinger (LS) equation with an ultraviolet cutoff $\Lambda $.
Currently, this prescription (the so-call Weinberg power counting (WPC)) has
been carried out to next-to-next-to-next-to leading order (N$^{3}$LO)%
\footnote{%
Note that the order here is defined based on the pion-exchange (long-range)
part of the potential, which does not necessary equal to the order at the
final NN amplitude.}, and has became the standard of many conventional
calculations\cite{EM03,Ep05,Ep12}. However, since Weinberg prescription
only applies power counting to the potential level, the systematic control
of the theory could be lost in the final amplitude.

\section{Renormalization group analysis}

One way to check whether a proposed scheme is under control is to perform
the RG-analysis. RG-analysis carried out at leading order (LO), up to
next-to-next-to leading order (NNLO) and N$^{3}$LO based on WPC indicate
that the conventional implementation of WPC fails to fulfill the RG
requirement once the ultraviolet cutoff of the iteration $\Lambda >1$ GeV%
\cite{NTvK05,Ya09A,Ya09B,ZE12}. Since the chiral expansion is established in
powers of $Q/M_{hi}$\footnote{%
Here $Q\equiv (p,m_{\pi })$, $p$ the NN c.m. momentum, $m_{\pi }$ the pion
mass, and the breakdown scale $M_{hi}$ is nominally $m_{\rho }\sim 4\pi
f_{\pi }$.}, some authors\cite{EG09} have questioned whether the theory is valid once intermediate states have
 $p\sim \Lambda >M_{hi}$. Thus, it makes
little sense to perform RG-analysis for $\Lambda >1$ GeV, even the final
on-shell $Q<<M_{hi}$.\ 

A second point of view\cite{Bi09,BY1,BY2,BY3,B,PV1,PV2} takes the final
amplitude as a partial sum of the (infinitely many) diagrams, then under the
assumption that a reasonable separation of scales exists\footnote{%
In the case where there is no reasonable separation of scales, EFT is
impossible.}, in any EFT one should be able to organize those diagrams in a
systematic way to absorb the unimportant physics into contact term(s) order
by order after a proper renormalization. Thus, as long as $Q<<M_{hi}$, the
impact of high-energy physics (which is well-represented by the contact
term(s)) in the final amplitude should reduce as the increase of $\Lambda $,
since the contribution from physics haven't been integrated out (i.e., from $%
\Lambda $ to $\infty $) becomes smaller and smaller.

The answer of the above in-debating issue actually depends on how the
diagrams are organized. It was shown that due to the fine-tuning of low
energy constants and a Wigner bond-like effect\cite{Wigner}, once a cutoff $%
\Lambda >1$ GeV is adopted the renormalization is effectively dominated by
one contact term under the WPC scheme\cite{ZE12}. Moreover, a full-iteration
of some type of irreducible two-pion-exchange diagrams could result in a
pole-like structure\cite{Baru}. Therefore, if one insists to build a NN
potential based on $\chi $EFT and utilizes it later in a conventional way
(e.g., inserts it as a potential in Schrodinger or LS equation), he or she
needs to stay in $500<\Lambda <1000$ MeV. The consequence is that a full
RG-based analysis becomes inapplicable.

To allow a full RG-analysis, one must give up treating the whole chiral
potential non-perturbatively. In other words, for the NN case there exists
no \textquotedblleft ideal potential" (in the traditional sense) to be
extracted or derived. Some parts of the diagrams have to be included
perturbatively. Recent works\cite{Bi09, BY1,BY2,BY3,PV1,PV2} which treat the
subleading chiral potentials in the framework of
Distorted-Wave-Born-Approximation enable a full RG- and power counting
analysis. Once the $\Lambda -$dependence is under well-control, the
estimation of the theoretical error becomes much easier, i.e., the error is
given by $O(Q^{n+1}/M_{hi}^{n+1}$) up to order-n in the new power counting
scheme.

We must point out that the lacks of a RG-analysis cannot rule out the
possibility that WPC under the specified range of $\Lambda $\ could generate
final amplitudes which has the correct power counting, but there is no way to
check this so far. On the other hand, a RG-correct scheme could converge too
slowly to be useful. Thus, before the full implementation to few- and
many-body calculations, one cannot determine the superiority of either
scheme. Nevertheless, it is of importance to start with a scheme which
allows a full RG-analysis first, then check the power counting step by step
to build the theory on a more solid ground.

\section{New Power counting and Future task}

\begin{figure}[tbp]
\begin{center}
\includegraphics[width=8cm]{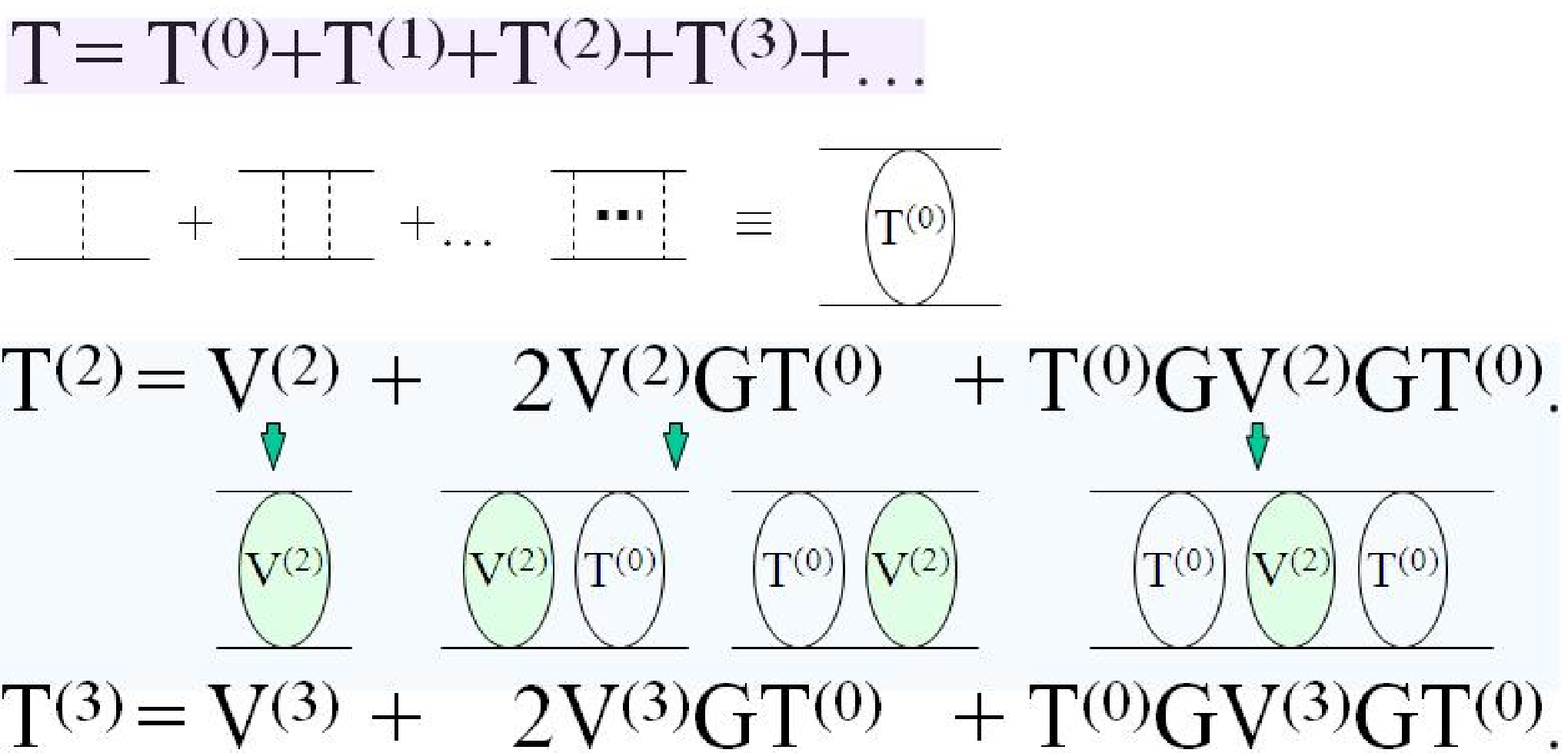}
\end{center}
\par
\vspace{3mm}
\caption{Diagrammatic representation of the new power counting in the case
where the $O(Q)$ contribution is absent. Here T, G and V denotes the T-matrix,
propagator and chiral potential. The order of T and V is indicated in the superscript.}
\label{fig-z}
\end{figure}

\bigskip The new power counting developed so far \cite{BY1,BY2,BY3} can be
summarized as:

1. The LO potential needs to be iterated to all order\footnote{%
At least for those spin-triplet and $l\leq 1$ partial-waves.}, and all
subleading chiral potentials are included perturbatively as represented
diagrammatically in Fig.~\ref{fig-z}.

2. The contact terms are determined by RG-analysis to guarantee the correct
RG-behavior. As a general rule, for potentials which are singular and
attractive at LO (i.e., $V_{LO}(r\rightarrow 0)\approx -\frac{1}{r^{n}}$
with $n\geq 3$), all contact terms need to be promoted one order earlier
with respect to WPC.

Phase shifts from the above new scheme are evaluated up to NNLO, and the
agreement with the Nijmegen phase-shift analysis~\cite{St93} is
comparable to those from WPC at the same order.

Furture tasks such as refining power counting with Lepage plot%
\cite{Le97} or similar techniques, including Delta (1232) contribution,
and deciding the power counting for high partial-waves ($l\geq 2$) are under
investigation.

\section*{Acknowledgments}
This work is supported by
the US NSF under grant PHYS-0854912, the MIUR grant PRIN-2009TWL3MX and US DOE under contract No.
DE-FG02-04ER41338, DE-AC05-06OR2317, E-AC05-06OR23177 and DE-FG02-93ER40756.

\end{document}